\documentclass[aps,prb,twocolumn,groupedaddress,showpacs,showkeys,amsmath,amssymb]{revtex4}

\usepackage{amsfonts}
\usepackage{amssymb,amsmath}
\usepackage{graphicx}
\usepackage{dcolumn}
\usepackage{bm}

\large\normalsize

\begin{document}

\title
{Regularly alternating spin--$\frac{1}{2}$ anisotropic $XY$ chains:
\\
The ground--state and thermodynamic properties}

\author
{Oleg Derzhko$^{1,2}$,
Johannes Richter$^3$,
Taras Krokhmalskii$^1$,
and
Oles' Zaburannyi$^1$}

\affiliation
{$^1$Institute for Condensed Matter Physics,
National Academy of Sciences of Ukraine,
\\
1 Svientsitskii Street, L'viv--11, 79011, Ukraine
\\
$^2$Chair of Theoretical Physics,
Ivan Franko National University of L'viv,
\\
12 Drahomanov Street, L'viv--5, 79005, Ukraine
\\
$^3$Institut f\"{u}r Theoretische Physik,
Universit\"{a}t Magdeburg,
\\
P.O. Box 4120, D--39016 Magdeburg, Germany}

\date
{\today}



\begin{abstract}
Using the Jordan--Wigner transformation and continued fractions
we calculate rigorously
the thermodynamic quantities
for the spin--$\frac{1}{2}$ transverse Ising chain
with periodically varying
intersite interactions and/or on--site fields.
We consider in detail the properties of the chains
having a period of the transverse field modulation equal to 3.
The regularly alternating transverse Ising chain
exhibits several quantum phase transition points,
where the number of transition points
for a given period of alternation
strongly depends on
the specific set of the Hamiltonian parameters.
The critical behavior in most cases is the same as for the uniform chain.
However,
for certain sets of the Hamiltonian parameters the critical behavior may be
changed and
weak singularities in the ground--state quantities appear.
Due to the regular alternation of the Hamiltonian parameters
the transverse Ising chain
may exhibit plateau--like steps
in the zero--temperature dependence
of the transverse magnetization vs. transverse field
and many--peak temperature profiles of the specific heat.
We compare the ground--state properties
of regularly alternating transverse Ising and transverse $XX$ chains
and of regularly alternating quantum and classical chains.

Making use of the corresponding unitary transformations
we extend the elaborated approach
to the study of thermodynamics
of regularly alternating spin--$\frac{1}{2}$ anisotropic $XY$ chains
without field.
We use the exact expression for the ground--state energy
of such a chain of period 2
to discuss how the exchange interaction anisotropy
destroys the spin--Peierls dimerized phase.
\end{abstract}

\maketitle

\section{Introductory remarks}
\label{s1}
\setcounter{equation}{0}

The spin--$\frac{1}{2}$ Ising chain in a transverse field
(transverse Ising chain)
is known as the simplest model
in the quantum theory of magnetism.
It can be viewed
as the 1D spin--$\frac{1}{2}$ anisotropic $XY$ model
in a transverse ($z$) field
with extremely anisotropic exchange interaction.
By means of the Jordan--Wigner transformation
it can be reduced
to a 1D model of noninteracting spinless fermions\cite{001,002,003,004}.
As a result the transverse Ising chain appeared to be an easy case\cite{005}
and a lot of studies on that model
have emerged up till now.
After the properties of the basic skeleton model
were understood
various modifications were introduced into the model
and the effects of the introduced changes were examined.
For example,
an analysis of the critical behavior
of the chain
with an aperiodic sequence of interactions
was performed in Ref. \onlinecite{006},
an extensive real--space renormalization--group treatment
of the random chain
was reported in Ref. \onlinecite{007},
a renormalization--group study
of the aperiodic chain
was presented in Ref. \onlinecite{008}.
It should be remarked, however,
that the simpler case
of the {\em regularly inhomogeneous}
spin--$\frac{1}{2}$ transverse Ising chain
(in which
the exchange interactions between the nearest sites
and/or the on--site transverse fields
vary regularly along the chain
with a finite period $p$)
still contains enough not explored properties
which deserve to be discussed.
Moreover,
the thermodynamic quantities of such a system
can be derived {\em rigorously analytically}
exploiting the fermionic representation and continued fractions.

The thermodynamic properties
of the regularly alternating anisotropic $XY$ chain
in a transverse field
of period 2
were considered in Refs. \onlinecite{009,010}
(see also Ref. \onlinecite{011}
where a model without field was investigated).
The elaborated general approach
for calculation of thermodynamic quantities\cite{010}
becomes rather tedious if $p>2$
and the properties of chains of larger periods of alternation
were not discussed.
Other papers\cite{012,013}
are devoted to the 1D anisotropic $XY$ model on superlattices,
which can be viewed as particular cases
of a regularly alternating anisotropic $XY$ chain.
In Ref. \onlinecite{012} the transfer matrix method
was applied
to get the excitation spectrum of the Hamiltonian,
being a quadratic form
of creation and annihilation Bose or Fermi operators,
on a 1D superlattice.
(This fermionic system is related
to the 1D spin--$\frac{1}{2}$ transverse anisotropic $XY$ model
on a superlattice.)
In Ref. \onlinecite{013}
a version of the approach suggested in Ref. \onlinecite{010}
was applied to  superlattices.
Considering as an example the ground--state dependences
of the transverse magnetization vs. transverse field
and
of the static transverse susceptibility vs. transverse field
(which were examined numerically
for an anisotropic $XY$ chain of period 4)
the authors of Ref. \onlinecite{013} observed
that these quantities behave differently
than for the isotropic $XY$ model. 
Contrary to the case of isotropic $XY$ model, 
for the  anisotropic $XY$ model
the number of the critical fields
at which the susceptibility becomes singular
strongly depends on the specific values of intersite interaction parameters.
The quantum critical points
in the anisotropic $XY$ chains in a transverse field
with periodically varying intersite interactions
(having periods 2 and 3)
were determined using the transfer matrix method in Ref. \onlinecite{014}.
It was found
that for periodic chains the number of quantum phase transition points
may increase
and its actual value
depends not only on the period of modulation
but also on the strengths of anisotropy and modulation of exchange interactions.
Let us also mention here a paper
discussing the energy gap vanishing
in the dimerized (i.e., period 2) anisotropic $XY$ chain without field\cite{015}
and a recent paper\cite{016}
which contains such an analysis for a nonzero transverse field.

In the present paper
we  have obtained a number of new results for regularly alternating
spin--$\frac{1}{2}$ anisotropic $XY$ chains
exploiting
a systematic method
for the calculation of the thermodynamic quantities
not used in the references cited above.
This approach is based on exploiting continued fractions\cite{017}
and seems to be a natural and convenient language
for describing regularly alternating chains
(Section \ref{s2}).
Considering the chains of period 3
we examine the generic effects
induced by regular alternation.
We discuss, in particular,
the effect of regular alternation
on the energy gap
(Sections \ref{s2} and \ref{s3}),
the zero--temperature dependences
of the transverse magnetization vs. transverse field
and
of the static transverse susceptibility vs. transverse field,
and
the temperature dependence of the specific heat
(Section \ref{s3}).
These rigorous analytical results
completed by numerical calculations of the spin correlation functions
(Section \ref{s3})
demonstrate the effect
of the regular alternation of  Hamiltonian parameters
on the quantum phase transition
inherent in the spin--$\frac{1}{2}$ transverse Ising chain.
Although
in most cases
the critical behavior remains like in the uniform chain case,
for certain sets of the Hamiltonian parameters
weak singularities in the ground--state quantities may appear.
We compare the results
for the transverse Ising chains
with the corresponding ones
for the isotropic $XY$ chains in a transverse field
(transverse $XX$ chains);
moreover,
we also compare the ground--state properties
of the quantum and classical
regularly alternating transverse Ising/$XX$ chains
(Section \ref{s3}).

The results obtained exploiting the continued fraction approach
can be used
to examine the thermodynamics
of the regularly alternating
spin--$\frac{1}{2}$ anisotropic $XY$ chain without field
since the latter model
is related to
a system of two spin--$\frac{1}{2}$ transverse Ising chains
through
certain unitary transformations.
We use the exact expression for the ground--state energy
of the anisotropic $XY$ chain without field
of period 2
to demonstrate
the effects of anisotropy of exchange interaction
on the spin--Peierls dimerization
(Section \ref{s4}).

We end up this Section
introducing notations
and making some symmetry remarks.
We consider $N\to\infty$ spins $\frac{1}{2}$ on a ring
governed by the Hamiltonian
\begin{eqnarray}
\label{1.01}
H=\sum_{n=1}^{N}\Omega_ns_n^z
+\sum_{n=1}^{N}2I_n s^x_ns^x_{n+1}
\end{eqnarray}
(with
$s_{N+1}^{\alpha}=s_1^{\alpha}$
($I_N=0$)
for periodic (open) boundary conditions).
Here $I_n$ is the (Ising) exchange interaction
between the nearest sites $n$ and $n+1$
and
$\Omega_{n}$ is the transverse field at the site $n$.
We assume that these quantities
vary regularly along the chain with period $p$,
i.e.,
the sequence of parameters in (\ref{1.01}) is
$\Omega_{1}I_1\Omega_{2}I_2\ldots\Omega_{p}I_p
\Omega_{1}I_1\Omega_{2}I_2\ldots\Omega_{p}I_p\ldots\;$.
Our goal is to examine
the thermodynamic properties of the spin model (\ref{1.01}).

Let us extend
the ``duality'' transformation\cite{018,003}
to the inhomogeneous case
(for the Ising chain in a random transverse field
such a transformation was discussed in Refs. \onlinecite{009,019}).
It can be easily proved
that the partition function
$Z={\mbox{Tr}}\exp\left(-\beta H\right)$
for two sequences of parameters
$\ldots\Omega_{n}I_n\ldots$
and
$\ldots I_{n-1}\Omega_{n}\ldots$
(or $\ldots I_n\Omega_{n+1}\ldots\;$)
is the same.
That means that the fields and the interactions
may be interchanged as
$\Omega_n\to I_{n-1}$
and
$I_n\to \Omega_n$
(or $\Omega_n\to I_n$
and
$I_n\to\Omega_{n+1}$)
remaining the partition function unchanged.
Really,
performing the unitary transformation
$U=\prod_{p=1}^{N-1}\exp\left({{\mbox{i}}\pi s_p^xs_{p+1}^y}\right)$
one finds that Eq. (\ref{1.01})
transforms
(with the accuracy to the end terms
not important for the thermodynamics)
into
\begin{eqnarray}
\label{1.02}
UH{U}^+
=\sum_{n=1}^{N}I_ns_{n+1}^z
+\sum_{n=1}^{N}2\Omega_n s^y_ns^y_{n+1}
\nonumber\\
=\sum_{n=1}^{N}I_ns_n^z
+\sum_{n=1}^{N}2\Omega_{n+1}s^y_ns^y_{n+1}
\end{eqnarray}
(to get the second equality
we have renumbered the sites
$n\to n-1$
which obviously does not change the thermodynamics).
As a result
$R^zUHU^+{R^z}^+$
with
$R^z=\prod_{q=1}^{N}
\exp\left({\mbox{i}}\frac{\pi}{2}s_q^z\right)$
(up to the end effects)
is again the transverse Ising chain,
however, with the exchange interaction
between the nearest sites $n$ and $n+1$
being equal to $\Omega_n$ (or $\Omega_{n+1}$)
and the transverse field at the site $n$
being equal to $I_{n-1}$ (or $I_n$).

We also recall
that the unitary transformation
$F_m=2s_m^x$
changes the sign of the transverse field at site $m$
in the Hamiltonian (\ref{1.01}),
whereas the unitary transformation
$B_m=\left(2s_1^z\right)\ldots\left(2s_m^z\right)$
changes the sign of the exchange interaction between the sites $m$ and $m+1$
in the Hamiltonian (\ref{1.01}).
The symmetry remarks
permit to reduce the range of parameters
for the study of the thermodynamics of the model.

Finally,
let us extend the relation between
the anisotropic $XY$ chain without field
and
the transverse Ising chain
(see, for example, Refs. \onlinecite{019,020})
to the inhomogeneous case.
Applying
the unitary transformation
$V=\prod_{p=1}^{N-1}\exp\left({{\mbox{i}}\pi s_p^ys_{p+1}^z}\right)$
to the Hamiltonian
\begin{eqnarray}
\label{1.03}
H=\sum_{n=1}^{N}
\left(
2I^x_ns_n^xs^x_{n+1}+2I^y_n s^y_ns^y_{n+1}
\right)
\end{eqnarray}
one gets
\begin{eqnarray}
\label{1.04}
VHV^+
=\sum_{n=1}^{N}
\left(
2I^x_ns_n^zs^z_{n+2}+I^y_n s^x_{n+1}
\right)
\end{eqnarray}
(with the accuracy to the end terms).
This is the Hamiltonian of two independent chains.
Performing further in Eq. (\ref{1.04})
a $\frac{\pi}{2}$ rotation of all spins about the $y$--axis
one finds
that $R^yVHV^+{R^y}^+$,
$R^y=\prod_{q=1}^{N}\exp\left({\mbox{i}}\frac{\pi}{2}s_q^y\right)$
(up to the end effects)
is
the Hamiltonian of two independent transverse Ising chains
(in the notations used in Eq. (\ref{1.01})), where
each of $\frac{N}{2}$ sites
is defined by the sequences of parameters
$\ldots I_{n+1}^yI_{n+2}^xI_{n+3}^yI_{n+4}^x\ldots$
and
$\ldots I_n^yI_{n+1}^xI_{n+2}^yI_{n+3}^x\ldots\;$.
We shall use the discussed relation in Section \ref{s4}
to study the thermodynamic properties
of regularly alternating anisotropic $XY$ chains without field (\ref{1.03}).

\section{Continued fraction approach}
\label{s2}
\setcounter{equation}{0}

To derive the thermodynamic quantities of the spin model (\ref{1.01})
we first express the spin Hamiltonian in fermionic language
by applying the Jordan--Wigner transformation\cite{001,002,003,004,005}.
As a result
we arrive at
a model of spinless fermions on a ring
governed by the Hamiltonian
which can be transformed
into the diagonal form
\begin{eqnarray}
\label{2.01}
H=\sum_{k=1}^N\Lambda_k
\left(\eta_k^+\eta_k-\frac{1}{2}\right),
\\
\left\{\eta_k^+,\eta_q\right\}=\delta_{kq},
\;\;\;
\left\{\eta_k,\eta_q\right\}=\left\{\eta_k^+,\eta_q^+\right\}=0
\nonumber
\end{eqnarray}
after performing a linear canonical transformation.
The coefficients of the transformation
are determined from the following equations\cite{001,003,019,021}
\begin{eqnarray}
\label{2.02}
\Omega_{n-1}I_{n-1}\Phi_{k,n-1}
+\left(\Omega_n^2+I_{n-1}^2-\Lambda_k^2\right)\Phi_{kn}
\nonumber\\
+\Omega_nI_n\Phi_{k,n+1}=0;
\end{eqnarray}
\begin{eqnarray}
\label{2.03}
\Omega_{n}I_{n-1}\Psi_{k,n-1}
+\left(\Omega_n^2+I_{n}^2-\Lambda_k^2\right)\Psi_{kn}
\nonumber\\
+\Omega_{n+1}I_n\Psi_{k,n+1}=0.
\end{eqnarray}
Evidently,
we may obtain the thermodynamic quantities of the spin model (\ref{1.01})
having the density of states
\begin{eqnarray}
\label{2.04}
R(E^2)=\frac{1}{N}\sum_{k=1}^N\delta(E^2-\Lambda_k^2)
\end{eqnarray}
since due to Eq. (\ref{2.01})
the Helmholtz free energy per site is given by
\begin{eqnarray}
\label{2.05}
f=-\frac{2}{\beta}\int_0^{\infty}{\mbox{d}}E
ER(E^2)\ln\left(2\cosh\frac{\beta E}{2}\right).
\end{eqnarray}
As we shall see below
the density of states $R(E^2)$
(which contains the same information as a set of all $\Lambda_k$)
is easier to calculate
than
the values of $\Lambda_k$
or the coefficients $\Phi_{kn}$, $\Psi_{kn}$.

On the other hand,
we may exploit
Eqs. (\ref{2.02}), (\ref{2.03})
to obtain the desired density of states $R(E^2)$ (\ref{2.04}).
We note
that the three diagonal band set of equations (\ref{2.02}) (or (\ref{2.03}))
strongly resembles the one
describing displacements of particles
in a nonuniform harmonic chain with nearest neighbor interactions
and $R(E^2)$ (\ref{2.04}) plays the role of the distribution
of the squared phonon frequencies
(for a study of the phonon density of states in a linear nonuniform system
see, for example, Ref. \onlinecite{022}).
The set of equations (\ref{2.02}) (or (\ref{2.03}))
can be also viewed as the one
for determining  a wave function of (spinless) electron
in the 1D nonuniform tight--binding model.

To find the density of states $R(E^2)$ from
the set of equations (\ref{2.02}) (or (\ref{2.03}))
we use the standard Green function approach.
Consider, for example, Eq. (\ref{2.02}).
Let us introduce the Green functions
$G_{nm}=G_{nm}(E^2)$
which satisfy the set of equations
\begin{eqnarray}
\label{2.06}
\left(E^2-\Omega_n^2-I_{n-1}^2\right)G_{nm}
-\Omega_{n-1}I_{n-1}G_{n-1,m}
\nonumber\\
-\Omega_{n}I_{n}G_{n+1,m}
=\delta_{nm}.
\end{eqnarray}
Knowing the diagonal Green functions $G_{nn}=G_{nn}(E^2)$
we immediately find the density of states $R(E^2)$ (\ref{2.04})
through the relation
\begin{eqnarray}
\label{2.07}
R(E^2)=
\mp\frac{1}{\pi N}\sum_{n=1}^N
{\mbox{Im}}G_{nn}(E^2\pm {\mbox{i}}\epsilon),
\;
\epsilon\rightarrow+0.
\end{eqnarray}

Alternatively, $R(E^2)$ can be obtained
with the help of the Green functions
introduced on the basis of the set of equations
for coefficients $\Psi_{kn}$ (\ref{2.03}).
The set of equations for such Green functions
(like Eq. (\ref{2.06}))
corresponds to the unitary equivalent spin chain
(see (\ref{1.02}))
which exhibits the same thermodynamic properties.
Thus,
the resulting density of states $R(E^2)$
is the same.

Now we have to calculate the diagonal Green functions
$G_{nn}$
involved into Eq. (\ref{2.07}).
Let us use the continued fraction representation
for $G_{nn}$
that follows from (\ref{2.06})
\begin{eqnarray}
\label{2.08}
G_{nn}
=\frac{1}
{E^2-\Omega_{n}^2-I_{n-1}^2-\Delta_n^--\Delta_n^+},
\\
\Delta_n^-
=\frac{\Omega_{n-1}^2I_{n-1}^2}
{E^2-\Omega_{n-1}^2-I_{n-2}^2
-\frac{\Omega_{n-2}^2I_{n-2}^2}
{E^2-\Omega_{n-2}^2-I_{n-3}^2-_{\ddots}}},
\nonumber\\
\Delta_n^+
=\frac{\Omega_{n}^2I_{n}^2}
{E^2-\Omega_{n+1}^2-I_{n}^2
-\frac{\Omega_{n+1}^2I_{n+1}^2}
{E^2-\Omega_{n+2}^2-I_{n+1}^2-_{\ddots}}}.
\nonumber
\end{eqnarray}
(Note that the signs of exchange interactions and fields
are not important for the thermodynamic quantities
as it was noted above
and is explicitly seen from Eq. (\ref{2.08}).)
For any finite period of varying
$\Omega_n$ and $I_n$
the continued fractions in (\ref{2.08}) become periodic
(in the limit $N\to\infty$)
and can be easily calculated by solving quadratic equations.
As a result we get rigorous expressions
for the diagonal Green functions,
the density of states (\ref{2.07})
and the thermodynamic quantities (\ref{2.05})

of the periodically alternating spin chain (\ref{1.01}). For example,
on gets for the internal energy
$e=f+\beta\frac{\partial f}{\partial\beta}$,
for the entropy
$\frac{s}{k}=\beta^2\frac{\partial f}{\partial \beta}$
or for the specific heat
$\frac{c}{k}=-\beta\frac{\partial}{\partial\beta}\frac{s}{k}$.
Assuming that $\Omega_n=\Omega+\Delta\Omega_n$
one can also obtain
the transverse magnetization
$m^z=\frac{\partial f}{\partial \Omega}$
and the static transverse susceptibility
$\chi^z=\frac{\partial m^z}{\partial \Omega}$.

Following the procedure described above,
for the periodically alternating chains of period 2 and 3
we find the following result for $R(E^2)$
\begin{eqnarray}
\label{2.09}
R(E^2)=
\left\{
\begin{array}{ll}
\frac{1}{p\pi}
\frac{\vert {\cal{Z}}_{p-1}(E^2)\vert}
{\sqrt{{\cal{A}}_{2p}(E^2)}}, &
{\mbox{if}}\;\;\;
{\cal{A}}_{2p}(E^2)>0, \\
0, &
{\mbox{otherwise}},
\end{array}
\right.
\end{eqnarray}
where
${\cal{Z}}_{p-1}(E^2)$
and
${\cal{A}}_{2p}(E^2)
=-\prod_{j=1}^{2p}\left(E^2-a_j\right)$
are polynomials of $(p-1)$th and $(2p)$th orders,
respectively,
and $0\le a_1\le \ldots \le a_{2p}$
are the roots of ${\cal{A}}_{2p}(E^2)$.
Moreover,
\begin{eqnarray}
\label{2.10}
{\cal{Z}}_1(E^2)
=2E^2-\Omega_1^2-\Omega_2^2-I_1^2-I_2^2,
\nonumber\\
{\cal{A}}_4(E^2)
=
4\Omega_1^2\Omega_2^2I_1^2I_2^2
\nonumber\\
-\left(
E^4
-\left(\Omega_1^2+\Omega_2^2+I_1^2+I_2^2\right)E^2
+\Omega_1^2\Omega_2^2+I_1^2I_2^2
\right)^2;
\end{eqnarray}
\begin{eqnarray}
\label{2.11}
{\cal{Z}}_2(E^2)
=3E^4
-2\left(
\Omega_1^2+\Omega_2^2+\Omega_3^2+I_1^2+I_2^2+I_3^2
\right)E^2
\nonumber\\
+\Omega_1^2\Omega_2^2+\Omega_1^2I_2^2+I_1^2I_2^2
+\Omega_2^2\Omega_3^2+\Omega_2^2I_3^2+I_2^2I_3^2
\nonumber\\
+\Omega_3^2\Omega_1^2+\Omega_3^2I_1^2+I_3^2I_1^2,
\nonumber\\
{\cal{A}}_6(E^2)
=4\Omega_1^2\Omega_2^2\Omega_3^2I_1^2I_2^2I_3^2
\nonumber\\
-\left(
E^6
-\left(\Omega_1^2+\Omega_2^2+\Omega_3^2+I_1^2+I_2^2+I_3^2\right)E^4
\right.
\nonumber\\
\left.
+\left(
\Omega_1^2\Omega_2^2+\Omega_1^2I_2^2+I_1^2I_2^2
+\Omega_2^2\Omega_3^2+\Omega_2^2I_3^2+I_2^2I_3^2
\right.
\right.
\nonumber\\
\left.
\left.
+\Omega_3^2\Omega_1^2+\Omega_3^2I_1^2+I_3^2I_1^2
\right)E^2
-\Omega_1^2\Omega_2^2\Omega_3^2
-I_1^2I_2^2I_3^2
\right)^2.
\end{eqnarray}

Eqs. (\ref{2.09}), (\ref{2.10}), (\ref{2.11})
recover the result for the uniform chain
if $\Omega_n=\Omega$, $I_n=I$
as it should be.
The obtained density of states for $p=2$ (\ref{2.09}), (\ref{2.10})
can be compared with the exact calculation
for the anisotropic $XY$ chain in a transverse field
reported in Ref. \onlinecite{010}.
Such a spin chain is represented
by noninteracting spinless fermions
with the energies $\Lambda_{\pm}(q)$
given by Eq. (2.22) of that paper.
The density of states (\ref{2.04})
has the form
$R(E^2)
=\frac{1}{2\pi}\sum_{\nu=\pm}
\int_{-\frac{\pi}{2}}^{\frac{\pi}{2}}{\mbox{d}}q
\delta\left(E^2-\Lambda^2_\nu(q)\right)$
and
for the transverse Ising chain of period 2
after a simple integration it transforms into (\ref{2.09}), (\ref{2.10}).

\begin{figure}[t]
\vspace{-10mm}
\epsfxsize=6.0cm
\center{\epsffile{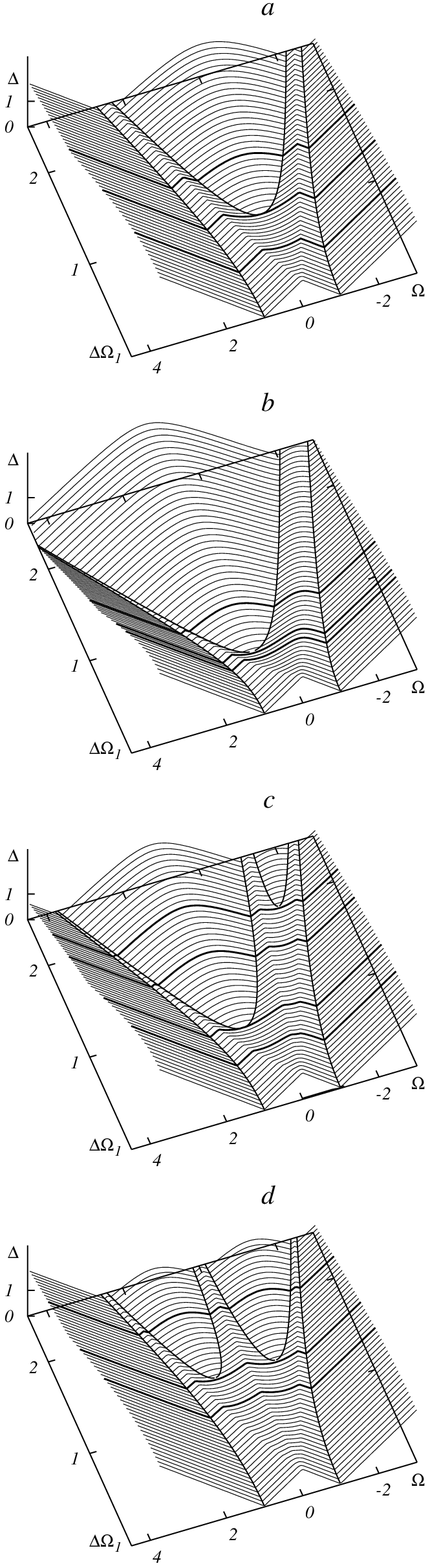}}
\caption
{The energy gap $\Delta$ vs. transverse field $\Omega$
for transverse Ising chains
of period 2 (panel a) and 3 (panels b -- d)
($\Omega_{n}=\Omega+\Delta\Omega_{n}$,
$\sum_{n=1}^p\Delta\Omega_n=0$,
$I_{n}=1$;
$\Delta\Omega_2=\Delta\Omega_1$ (b),
$\Delta\Omega_2=\frac{1}{2}\Delta\Omega_1$ (c),
$\Delta\Omega_2=-\Delta\Omega_1$ (d)).
The bold curves $\Delta$ vs. $\Omega$
correspond to the following values of parameters:
$\Delta\Omega_1=0.5,\;1,\;1.5$ (panel a),
$\Delta\Omega_1=0.5$,
$\Delta\Omega_1\approx 0.630$,
$\Delta\Omega_1=1$ (panel b),
$\Delta\Omega_1=0.5$,
$\Delta\Omega_1\approx 0.848$,
$\Delta\Omega_1=1.5$,
$\Delta\Omega_1\approx 1.921$ (panel c),
$\Delta\Omega_1=1$,
$\Delta\Omega_1\approx 1.375$,
$\Delta\Omega_1=2$ (panel d).
The bold curves in the plane
$\Omega$--$\Delta\Omega_1$
indicate
the values of parameters
which yield the zero--energy gap, $\Delta=0$.
\label{fr01}}
\end{figure}

The density of states $R(E^2)$ (\ref{2.04}) yields
valuable information
about the spectral properties of the Hamiltonian (\ref{1.01}).
Thus, the gap $\Delta$
in the energy spectrum of the spin chain
is given by the square root of the smallest root $a_1$
of the polynomial ${\cal{A}}_{2p}(E^2)$.
In Fig. \ref{fr01}
we display the dependence of the energy gap on the transverse field\footnote
{Hereafter we fix
$\Delta\Omega_n$
($\Omega_n=\Omega+\Delta\Omega_n$, $\sum_n\Delta\Omega_n=0$)
and
$I_n$
and assume $\Omega$ to be a free parameter.
As a result we consider
the changes in different properties
as the transverse field (and temperature) varies.
Obviously,
while fixing
$\Omega_n$
and
$\Delta I_n$
($I_n=I+\Delta I_n$, $\sum_n\Delta I_n=0$)
and assuming $I$ to be a free parameter
(in such a case,
e.g.,
the quantum phase transition is driven
by tuning the exchange interaction $I$
rather than the transverse field $\Omega$)
we recover the ``violated'' symmetry
between transverse field and exchange interaction.}
for some chains of period 2 and 3.
The vanishing gap indicates quantum phase transition points\cite{004}.
As can be seen from the data reported in Fig. \ref{fr01}
the number of such quantum phase transition points
for a given period of alternation
is strongly parameter--dependent.
The chains of period 2 may become gapless
either at one, two, three, or four values of the transverse field,
whereas the chains of period 3 may become gapless
either at one, two, three, four, five, or six values of the transverse field
depending on the specific set of the Hamiltonian parameters.
The condition for the vanishing gap
follows from
${\cal{A}}_4(0)=0$
(\ref{2.10})
(${\cal{A}}_6(0)=0$
(\ref{2.11}))
and for the chains of period 2 (3) it reads
$\Omega_1\Omega_2=\pm I_1I_2$
($\Omega_1\Omega_2\Omega_3=\pm I_1I_2I_3$).
In fact
we have rederived with the help of continued fractions
the long known condition for the existence of the zero--energy excitations
in the inhomogeneous spin--$\frac{1}{2}$ transverse Ising chain\cite{023}
which in our notations has the form
\begin{eqnarray}
\label{2.12}
\Omega_1\Omega_2\ldots\Omega_N=\pm I_1I_2\ldots I_N.
\end{eqnarray}
(Notice,
that Eq. (6) of Ref. \onlinecite{023} does not contain two signs;
the minus sign follows from the symmetry arguments
after performing simple rotations of spin axes.
It is important, as will be seen below,
to have two signs in (\ref{2.12}).)
Obviously, for periodic chains we have
the products of only $p$ multipliers
in the l.h.s. and r.h.s. of Eq.  (\ref{2.12}).

For a chain of period 2 with a uniform transverse field
Eq. (\ref{2.12}) yields
either one critical field $\Omega^\star=0$
if either $I_1$ or $I_2$ (or both) equals to zero
or two critical fields
$\Omega^{\star}=\pm\sqrt{\vert I_1I_2\vert}$.
If  the transverse field becomes regularly varying,
$\Omega_{1,2}=\Omega\pm\Delta\Omega$,
$\Delta\Omega>0$,
there may be either two critical fields
$\Omega^{\star}=\pm\sqrt{\Delta\Omega^2+\vert I_1I_2\vert}$
if
$\Delta\Omega<\sqrt{\vert I_1I_2\vert}$,
or three critical fields
$\Omega^{\star}=\left\{\pm\sqrt{2\vert I_1I_2\vert}, 0\right\}$
if
$\Delta\Omega=\sqrt{\vert I_1I_2\vert}$,
or four critical fields
$\Omega^{\star}=\pm\sqrt{\Delta\Omega^2\pm\vert I_1I_2\vert}$
if
$\Delta\Omega>\sqrt{\vert I_1I_2\vert}$
(see Fig. \ref{fr01}a).
As a result
a chain of period 2 with
$\Delta\Omega<\sqrt{\vert I_1I_2\vert}$,
as $\Omega$ varies, exhibits two phases:
the Ising phase
(for $\vert\Omega\vert<\sqrt{\Delta\Omega^2+\vert I_1I_2\vert}$)
and the paramagnetic phase
(for $\vert\Omega\vert>\sqrt{\Delta\Omega^2+\vert I_1I_2\vert}$).
A chain of period 2 with
$\Delta\Omega=\sqrt{\vert I_1I_2\vert}$,
as $\Omega$ varies, also exhibits two phases:
the Ising phase
(for $0<\vert \Omega\vert <\sqrt{2\vert I_1I_2\vert}$)
and
the paramagnetic phase
(for $\vert \Omega\vert >\sqrt{2\vert I_1I_2\vert}$);
moreover,
in the Ising phase at $\Omega=\Omega^\star=0$
the system exhibits a weak singularity in the ground--state quantities (see below).
A chain of period 2 with
$\Delta\Omega>\sqrt{\vert I_1I_2\vert}$,
as $\Omega$ varies, exhibits three phases:
the low--field paramagnetic phase
(for $\vert\Omega\vert<\sqrt{\Delta\Omega^2-\vert I_1I_2\vert}$),
the Ising phase
(for $\sqrt{\Delta\Omega^2-\vert I_1I_2\vert}<\vert\Omega\vert
<\sqrt{\Delta\Omega^2+\vert I_1I_2\vert}$),
and
the high--field paramagnetic phase
(for $\vert\Omega\vert>\sqrt{\Delta\Omega^2+\vert I_1I_2\vert}$).
A motivation to give such names to different phases
follows from the behavior of the Ising magnetization $m^x$
to be discussed below in Section \ref{s3}.

\begin{figure}[t]
\epsfxsize=8.0cm
\center{\epsffile{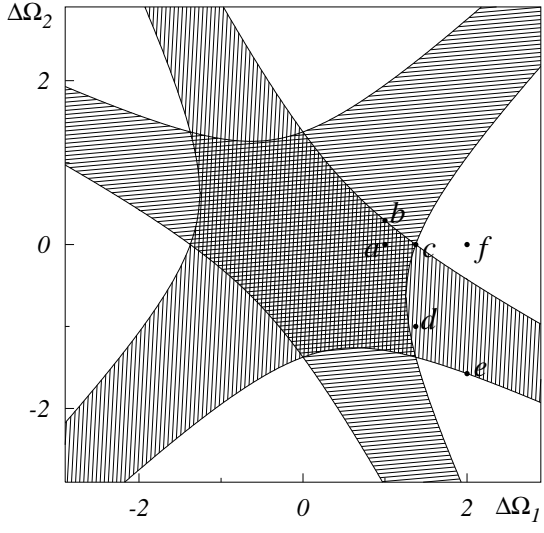}}
\caption
{Phase diagram of the transverse Ising chain of period 3
with $\vert I_1I_2I_3 \vert =1$,
$\Omega_{1,2,3}=\Omega+\Delta\Omega_{1,2,3}$,
$\Delta\Omega_1+\Delta\Omega_2+\Delta\Omega_3=0$.
As $\Omega$ varies the energy gap vanishes
two/four/six times
if the set of parameters is
in the dark/gray/light region.
The sets of parameters
denoted by a, b, c, d, e, f
are used below to illustrate the dependence on $\Omega$
of the ground--state Ising magnetization
(Fig. \ref{fr05})
and the low--temperature specific heat
(Fig. \ref{fr07}).
\label{fr02}}
\end{figure}

For a chain of period 3
($\Omega_{1,2,3}
=\Omega+\Delta\Omega_{1,2,3}$,
$\Delta\Omega_1+\Delta\Omega_2+\Delta\Omega_3=0$)
the critical fields follow from two cubic equations
\begin{eqnarray}
\label{2.13}
\left(\Omega^\star+\Delta\Omega_1\right)
\left(\Omega^\star+\Delta\Omega_2\right)
\left(\Omega^\star+\Delta\Omega_3\right)
\pm I_1I_2I_3=0
\end{eqnarray}
each of which may have
either one real solution
or three real solutions.
In Fig. \ref{fr02}
we display
the regions in $\Delta\Omega_1$--$\Delta\Omega_2$ plane
for the set of parameters
of the transverse Ising chains
with $\vert I_1I_2I_3\vert=1$
which yield two (dark region), four (gray region), or six (light region)
values of the critical field.
For the set of parameters at the boundary between dark and gray
(gray and light) regions
there are three (five) critical fields;
for the set of parameters where dark, gray and light regions meet
there are four critical fields.
The behavior of the energy gap for all cases
can be seen in Figs. \ref{fr01}b -- \ref{fr01}d.
As a result the chain of period 3
depending on a relation between
$\Delta\Omega_1$,
$\Delta\Omega_2$,
$\Delta\Omega_3$
may exhibit either two phases
(the Ising and paramagnetic phases),
or four phases
(two Ising and two paramagnetic phases),
or six phases
(three Ising and three paramagnetic phases).
Moreover,
weak singularities in the Ising phases may occur.

\section{The ground--state and thermodynamic properties}
\label{s3}
\setcounter{equation}{0}

\subsection{The ground--state magnetic properties:
Transverse Ising chain vs. transverse $XX$ chain}

\begin{figure}[t]
\epsfxsize=15.0cm
\center{\epsffile{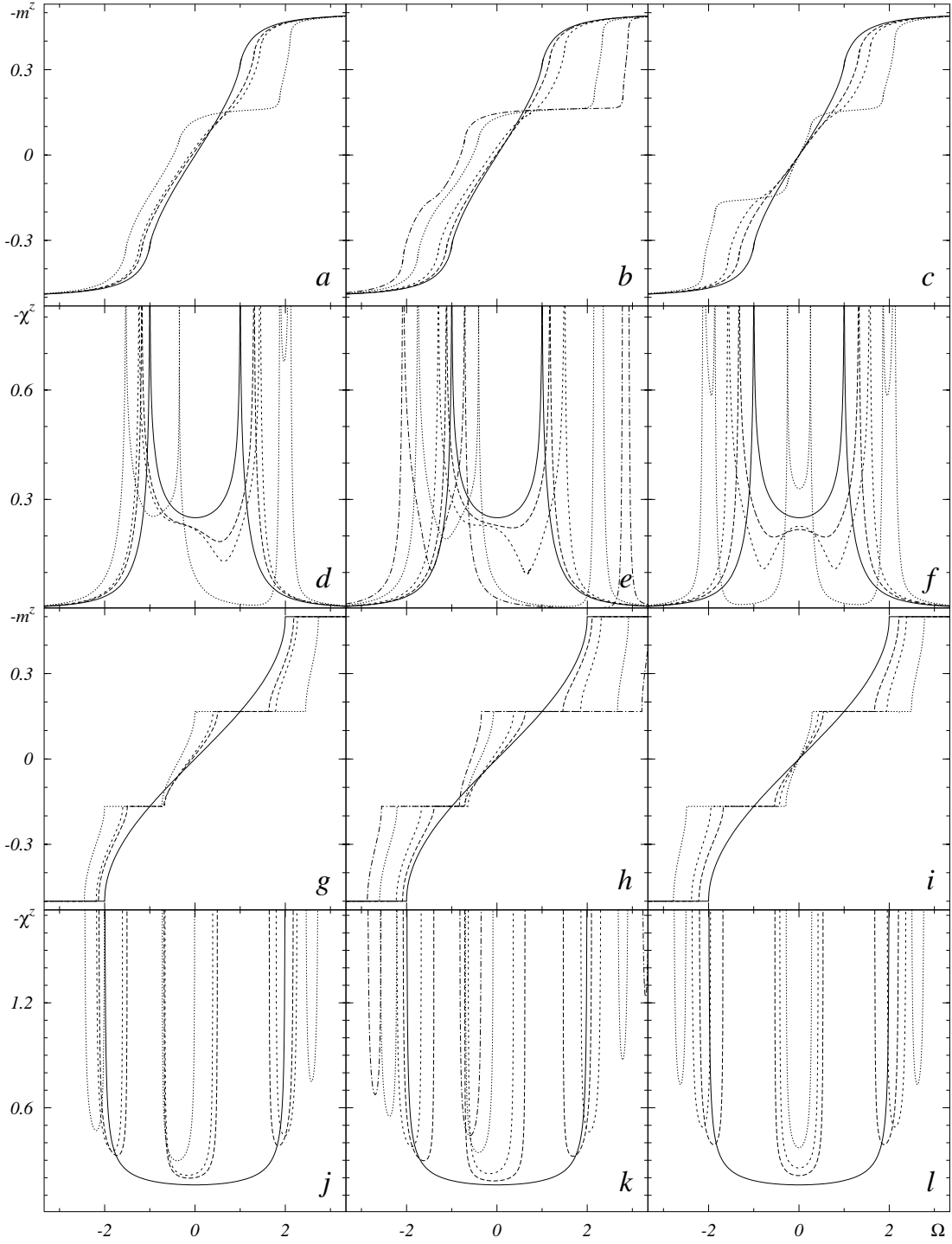}}
\caption
{The ground--state transverse magnetization (a, b, c, g, h, i)
and static transverse susceptibility (d, e, f, j, k, l) curves
for transverse Ising (a -- f)
and
transverse $XX$ (g -- l)
chains of period $p=3$.
$I_{1,2,3}=1$,
$\Omega_{1,2,3}=\Omega+\Delta\Omega_{1,2,3}$,
$\Delta\Omega_1+\Delta\Omega_2+\Delta\Omega_3=0$;
a, d, g, j:
$\Delta\Omega_2=\Delta\Omega_1$,
$\Delta\Omega_1=0$ (solid curves),
$\Delta\Omega_1=0.5$ (long--dashed curves),
$\Delta\Omega_1=0.6$ (short--dashed curves),
$\Delta\Omega_1=1$ (dotted curves);
b, e, h, k:
$\Delta\Omega_2=\frac{1}{2}\Delta\Omega_1$,
$\Delta\Omega_1=0$ (solid curves),
$\Delta\Omega_1=0.5$ (long--dashed curves),
$\Delta\Omega_1=0.85$ (short--dashed curves),
$\Delta\Omega_1=1.5$ (dotted curves),
$\Delta\Omega_1=1.9$ (long--dashed--dotted curves);
c, f, i, l:
$\Delta\Omega_2=-\Delta\Omega_1$,
$\Delta\Omega_1=0$ (solid curves),
$\Delta\Omega_1=1$ (long--dashed curves),
$\Delta\Omega_1=1.35$ (short--dashed curves),
$\Delta\Omega_1=2$ (dotted curves).
\label{fr03}}
\end{figure}

The transverse magnetization and the static transverse susceptibility
for a regularly alternating transverse Ising chain
can be obtained using continued fractions
as was explained in Section \ref{s2}.
Such results for some typical chains of period 3
(which roughly correspond to the parameters
singled out in Figs. \ref{fr01}b -- \ref{fr01}d)
at zero temperature
are reported in Fig. \ref{fr03}.
Let us compare and contrast
the results for the magnetic properties
of the transverse Ising and the transverse $XX$ chains.

We start from the energy gap.
It is known that the uniform transverse Ising chain becomes gapless
at critical field $\Omega^\star=\pm \vert I\vert$.
The gap decays linearly while the transverse field
approaches the critical value,
$\Delta\sim\epsilon$,
$\epsilon=\vert\Omega-\Omega^\star\vert\to 0$.
The transverse $XX$ chain is gapless along the critical line
$-\vert I\vert \le \Omega \le \vert I\vert$.
The gap opens linearly while the value of transverse field
exceeds $\vert I\vert$.
If regular inhomogeneity is introduced
into the transverse $XX$ chain
the critical line splits into several parts;
the gaps open linearly as the transverse field runs out the critical lines\cite{024}.
On the contrary,
a regular inhomogeneity introduced into the transverse Ising chain
may either only shift the values of critical fields
or lead to the appearance of new critical points.
Moreover,
the gap decays either linearly,
$\Delta\sim\epsilon$,
or proportionally to the deviation from the  critical value squared,
$\Delta\sim\epsilon^2$,
as can be seen in Fig. \ref{fr01} (see also below).

The energy gap behavior determines
the zero--temperature transverse magnetization curves
for both chains.
Transverse $XX$ chains exhibit plateaus
which can be easily understood within the frames of fermionic picture.
Indeed, a regularly alternating transverse $XX$ chain
corresponds to a system of free fermions
with several energy bands
and the transverse field
plays the role of the chemical potential.
Transverse Ising chains do not exhibit plateaus,
however,
being in the paramagnetic phases exhibit plateau--like steps
(compare the curves
in Figs. \ref{fr03}a -- \ref{fr03}c
and
in Figs. \ref{fr01}b -- \ref{fr01}d).
In the Ising phases
the transverse magnetization shows a rapid change.
In the fermionic picture
a regularly alternating transverse Ising chain again corresponds to
a system of free fermions
with several energy bands,
however, the transverse field
does not play the role of the chemical potential any more.

The described behavior
of the transverse magnetization vs. transverse field
is accompanied
by the corresponding peculiarities in the behavior
of the static transverse susceptibility vs. transverse field.
Thus,
in the cases of the transverse $XX$ chain
the square--root singularities
indicate the gapless--to--gapped transitions
(Figs. \ref{fr03}j -- \ref{fr03}l).
In the case of the transverse Ising chain
a linear gap decay produces a logarithmic singularity
(Figs. \ref{fr03}d -- \ref{fr03}f),
whereas
for a decay proportional to the squared deviation
from the critical value
the static transverse susceptibility does not diverge
containing, however, a nonanalytical contribution
which causes a logarithmic singularity of its second derivative
(short--dashed and long--dashed--dotted curves
in Figs. \ref{fr03}d -- \ref{fr03}f).

To end up,
we emphasize
that for the regularly alternating transverse $XX$ chains
the number of peculiarities
(e.g.,
in the dependence $\chi^z$ vs. $\Omega$)
depends only on the period of alternation
and equals $2p$.
This is not the case
for the regularly alternating transverse Ising chains:
the number of peculiarities cannot exceed $2p$
but may be smaller;
the actual number of peculiarities and their type
essentially depends on the specific set of the Hamiltonian parameters.
Let us also underline a similarity of these results
with what has been found
for the anisotropic/isotropic $XY$ models
on 1D superlattices\cite{013,014,025}.

\subsection{The ground--state magnetic properties:
Quantum chain vs. classical chain}

To demonstrate the role of quantum effects
in the zero--temperature magnetization processes
we consider
the classical counterparts of
regularly alternating transverse Ising and transverse $XX$ chains
(some calculations of the thermodynamic quantities
of the uniform classical spin chains can be found in Ref. \onlinecite{026}).
The classical spin model
consists of classical spins (vectors)
${\bf {s}}=(s,\theta,\phi)$
($0\le\theta\le\pi$
and
$0\le\phi<2\pi$
are the spherical coordinates of the spin)
on a ring
which interact with each other and an external field
and are governed
either by the Hamiltonian
\begin{eqnarray}
\label{3.01}
H=\sum_{n=1}^N\Omega_n s\cos\theta_n
\nonumber\\
+\sum_{n=1}^N 2I_n s^2\sin\theta_n\sin\theta_{n+1}\cos\phi_n\cos\phi_{n+1}
\end{eqnarray}
(transverse Ising chain)
or by the Hamiltonian
\begin{eqnarray}
\label{3.02}
H=\sum_{n=1}^N\Omega_n s\cos\theta_n
\nonumber\\
+\sum_{n=1}^N 2I_n s^2\sin\theta_n\sin\theta_{n+1}\cos\left(\phi_n-\phi_{n+1}\right)
\end{eqnarray}
(transverse $XX$ chain).
In Eqs. (\ref{3.01}), (\ref{3.02})
$s$ is the value of the spin
which plays only a quantitative role
(further we put $s=\frac{1}{2}$)
and the sequence of parameters
for a regularly alternating chain of period $p$
is again
$I_1\Omega_1I_2\Omega_2\ldots I_p\Omega_p
I_1\Omega_1I_2\Omega_2\ldots I_p\Omega_p\ldots\;$.
In what follows we restrict ourselves to the case
$I_n=I$,
$\Omega_n=\Omega+\Delta\Omega_n$,
$\sum_n\Delta\Omega_n=0$
which has already been discussed in some detail above.
Our goal is to examine the effect of regular inhomogeneity
on the ground--state properties
of the classical transverse Ising and transverse $XX$ chains.

Consider at first the transverse Ising chain.
One can easily construct
the ground--state spin configuration
and
the corresponding ground--state energy ansatz.
According to (\ref{3.01})
to minimize the ground--state energy
one should place all spins in $xz$ plane
(i.e.,
$\phi_n=\phi_{n+1}=\ldots=0(\pi)$
if $I<0$
or
$\phi_n=\phi_{n+2}=\ldots=0(\pi)$,
$\phi_{n+1}=\phi_{n+3}=\ldots=\pi(0)$
if $I>0$).
Moreover,
the angles $\theta_n$ are determined to minimize
the sum
of the contribution coming from the interaction with the transverse field
and
of the contribution coming from the intersite interaction
taking into account the period of inhomogeneity.
Thus,
for the chain of period $p$,
an ansatz for the ground--state energy per site reads
\begin{eqnarray}
\label{3.03}
\frac{E(\theta_1,\ldots,\theta_p)}{N}
=
\frac{s}{p}\sum_{n=1}^p\Omega_n \cos\theta_n
\nonumber\\
-\frac{2\vert I\vert s^2}{p}
\sum_{n=1}^p\sin\theta_n\sin\theta_{n+1}
\end{eqnarray}
and the angles $\theta_n$ are determined from the set of equations
\begin{eqnarray}
\label{3.04}
\frac{\partial}{\partial\theta_n}
\frac{E(\theta_1,\ldots,\theta_p)}{N}=0,
\;\;\;
n=1,\ldots, p.
\end{eqnarray}
Substituting the solution of Eq. (\ref{3.04})
(which yields the lowest ground--state energy)
into Eq. (\ref{3.03})
we get the ground--state energy of the chain.
Now the ground--state on--site magnetizations
are given by
$m^z_n=s\cos\theta_n$,
$m^x_n=s\sin\theta_n\cos\phi_n$.
We can also find the ground--state on--site static transverse susceptibility
$\chi^z_n=\frac{\partial m^z_n}{\partial \Omega}$.

Let us turn to the transverse $XX$ chain (\ref{3.02}).
In the ground--state spin configuration
the spin components in $xy$ plane
are directed arbitrarily but coherently
at all sites
having the values
$\left\vert m^\perp_n \right\vert=s\sin\theta_n$
(i.e.,
$\phi_n=\phi_{n+1}=\ldots=\phi$
($\phi$ is an arbitrary angle)
if $I<0$
or
$\phi_{n}=\phi_{n+2}=\ldots=\phi$
($0\le\phi<\pi$),
$\phi_{n+1}=\phi_{n+3}=\ldots=\phi+\pi$
if $I>0$).
An ansatz for the ground--state energy per site
is again given by Eq. (\ref{3.03})
and the angles $\theta_n$
are determined from Eq. (\ref{3.04}).
Moreover,
$m^z_n=s\cos\theta_n$
and
$\chi^z_n=\frac{\partial m^z_n}{\partial \Omega}$.

\begin{figure}[t]
\epsfxsize=14.0cm
\vspace{-15mm}
\center{\epsffile{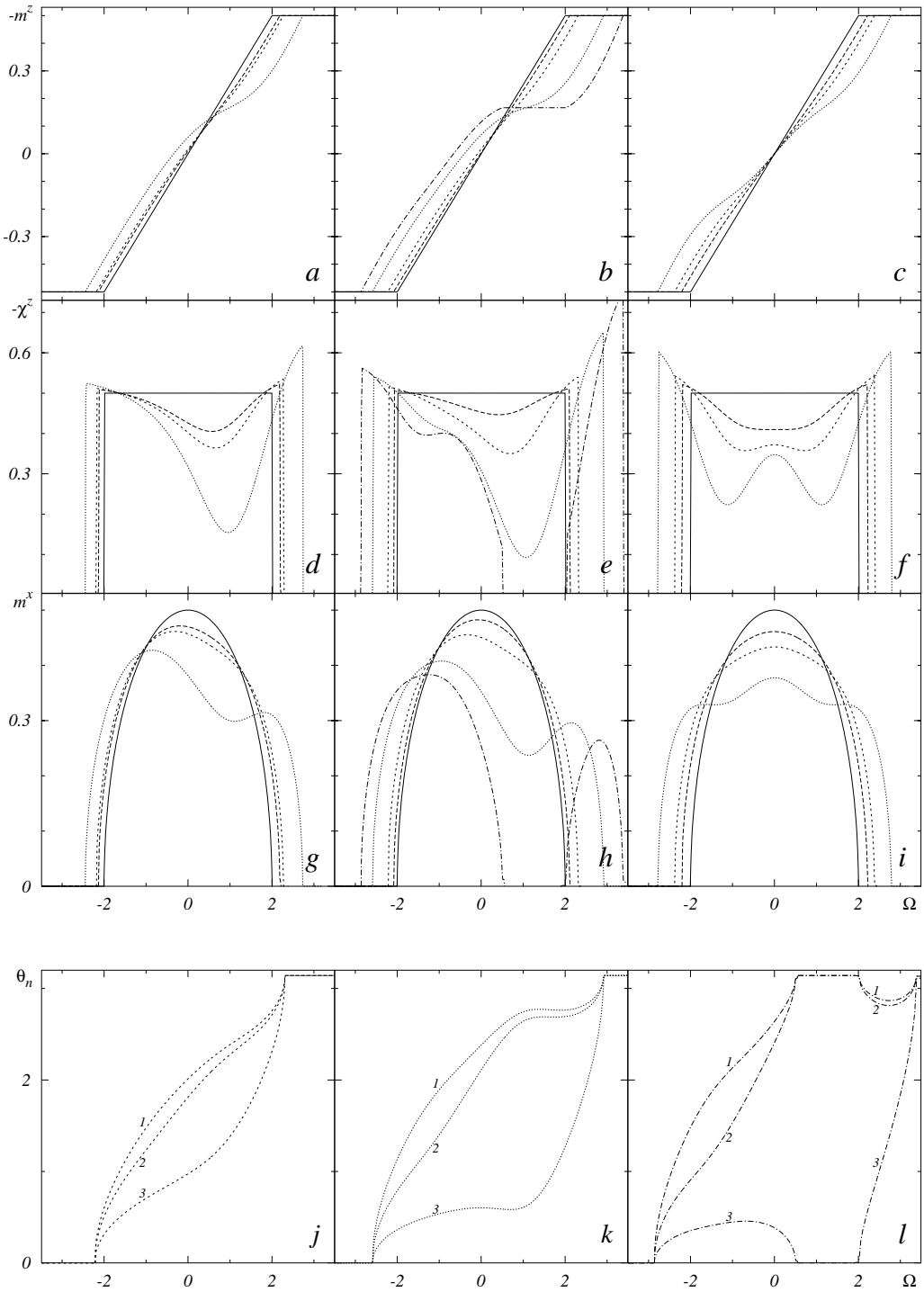}}
\caption
{The ground--state transverse magnetization (a, b, c),
Ising magnetization (g, h, i),
and static transverse susceptibility (d, e, f) curves
for the classical transverse Ising/$XX$ chains of period $p=3$.
$I_{1,2,3}=-1$,
$\Omega_{1,2,3}=\Omega+\Delta\Omega_{1,2,3}$,
$\Delta\Omega_1+\Delta\Omega_2+\Delta\Omega_3=0$;
a, d, g:
$\Delta\Omega_2=\Delta\Omega_1$,
$\Delta\Omega_1=0$ (solid curves),
$\Delta\Omega_1=0.5$ (long--dashed curves),
$\Delta\Omega_1=0.6$ (short--dashed curves),
$\Delta\Omega_1=1$ (dotted curves);
b, e, h:
$\Delta\Omega_2=\frac{1}{2}\Delta\Omega_1$,
$\Delta\Omega_1=0$ (solid curves),
$\Delta\Omega_1=0.5$ (long--dashed curves),
$\Delta\Omega_1=0.85$ (short--dashed curves),
$\Delta\Omega_1=1.5$ (dotted curves),
$\Delta\Omega_1=1.9$ (long--dashed--dotted curves);
c, f, i:
$\Delta\Omega_2=-\Delta\Omega_1$,
$\Delta\Omega_1=0$ (solid curves),
$\Delta\Omega_1=1$ (long--dashed curves),
$\Delta\Omega_1=1.35$ (short--dashed curves),
$\Delta\Omega_1=2$ (dotted curves).
We also show the ground--state spin configurations
$\theta_1$, $\theta_2$, $\theta_3$
(the corresponding curves are denoted by 1, 2, 3)
for the chains with
$\Delta\Omega_2=\frac{1}{2}\Delta\Omega_1$
(see panels b, e, h)
and
$\Delta\Omega_1=0.85$ (j),
$\Delta\Omega_1=1.5$ (k),
$\Delta\Omega_1=1.9$ (l)
as $\Omega$ varies.
\label{fr04}}
\end{figure}

For the chain of period 1
from (\ref{3.04})
one easily finds
$\theta=0$ if $\omega=\frac{\Omega}{4s\vert I\vert}<-1$,
$\theta=\arccos\left(-\omega\right)$
if $-1 \le\omega< 1$,
and
$\theta=\pi$
if $1 \le \omega$.
For the chain of period 2
from Eq. (\ref{3.04})
in addition to four obvious solutions
$\cos^2\theta_1=\cos^2\theta_2=1$
one gets one more solution
\begin{eqnarray}
\label{3.05}
\cos^2\theta_1
=\left(\omega+\delta\right)^2
\frac{1+\left(\omega-\delta\right)^2}{1+\left(\omega+\delta\right)^2},
\nonumber\\
\cos^2\theta_2
=\left(\omega-\delta\right)^2
\frac{1+\left(\omega+\delta\right)^2}{1+\left(\omega-\delta\right)^2};
\\
\delta=\frac{\Delta\Omega}{4s\vert I\vert}
\nonumber
\end{eqnarray}
if
$\vert\omega^2-\delta^2\vert \le 1$.
For the chain of period 3
Eq. (\ref{3.04}) has again obvious solutions
$\cos^2\theta_1=\cos^2\theta_2=\cos^2\theta_3=1$;
another solution
existing at a certain range of the transverse field
can be found numerically
(see Figs. \ref{fr04}j -- \ref{fr04}l).
The described analytical calculations
reproduce the results obtained earlier numerically
for some chains of periods 2 and 3
(dashed curves in Figs. 8a, 8b of Ref. \onlinecite{024}).

In Figs. \ref{fr04}a -- \ref{fr04}i
we display the obtained dependences
of the ground--state magnetizations $m^z$, $m^x$ and static transverse susceptibility $\chi^z$
on the transverse field
for several classical transverse Ising/$XX$ chains of period 3
(the results for corresponding quantum chains
are shown in Fig. \ref{fr03}).
In contrast to the quantum case,
the ground--state static transverse susceptibility $\chi^z$
of the classical chains
remains always finite as the transverse field $\Omega$ varies
and hence the classical chains do not exhibit
any ground--state phase transitions driven by $\Omega$.
However,
a regularly alternating classical chain
similarly to its quantum ($XX$) counterpart
may exhibit plateaus
in the ground--state dependence
transverse magnetization $m^z$ vs. transverse field $\Omega$
(compare long--dashed--dotted curves
in Fig. \ref{fr04}b (classical chain)
and Fig. \ref{fr03}h (quantum chain)
which have a plateau $-m^z=\frac{1}{6}$).
Obviously,
as $m^z$ remains constant with varying $\Omega$,
the static transverse susceptibility is zero
(long--dashed--dotted curves in Fig. \ref{fr04}e).
Moreover,
$m_n^x$ ($m_n^\perp$), $n=1,2,3$
in this region is zero
(Fig. \ref{fr04}h)
and the stable ground--state spin configuration is
$\theta_n=\theta_{n+1}=\pi$,
$\theta_{n+2}=0$
(see Fig. \ref{fr04}l).
The Ising magnetization $m^x$ decays as the system runs out the Ising phase
according to the power--law,
$m^x\sim\vert\Omega^\star-\Omega\vert^\beta$,
with $\beta=\frac{1}{2}$.

\subsection{Quantum phase transitions}

Let us have a closer look
at the quantum phase transitions in regularly alternating transverse Ising chains
discussing in some detail
the critical behavior.
For such a chain of period $p$
the quantum phase transition points are determined by Eq. (\ref{2.12}).
The effects of regular alternation on the number and position
of the quantum phase transition points
in the cases $p=2$ and $p=3$
were analyzed in Section \ref{s2}.
Although Eq. (\ref{2.12}) was found many years ago\cite{023}
it was not discussed in the context of the quantum phase transition theory.
In particular,
an important question {\em how} the gap vanishes
as the set of parameters becomes critical
was not considered in Ref. \onlinecite{023}.
Below we show that two types of critical behavior are possible:
one as it occurs for the second--order phase transition
(in Ehrenfest's sense)
and
another one as it occurs for a weaker singularity
(the fourth--order phase transition
in Ehrenfest's sense).
These findings are confirmed by numerical computations
of the two--site spin correlation functions.

First we analyze how the gap vanishes
as the set of parameters becomes critical
for
the case $p=2$
when
$\Omega^\star=\pm \sqrt{\Delta\Omega^2\pm \vert I_1I_2\vert}$.
Eq. (\ref{2.10}) yields
\begin{eqnarray}
\label{3.06}
\Delta^2(\Omega)
\approx
\frac{\left(\Omega^2-{\Omega^\star}^2\right)^2}
{2\left(\Omega^2+{\Delta\Omega}^2+\frac{I_1^2+I_2^2}{2}\right)}.
\end{eqnarray}
If $\vert\Omega-\Omega^{\star}\vert=\epsilon\to 0$
and $\Omega^\star\ne 0$
Eq. (\ref{3.06}) suggests
$\Delta^2(\Omega)\sim\epsilon^2$,
i.e.,
the energy gap vanishes linearly
(see Fig. \ref{fr01}a).
A linear decay of the energy gap
can be also seen in many cases in Figs. \ref{fr01}b -- \ref{fr01}d
for chains of period 3.
The linearly vanishing gap
corresponds to
the square--lattice Ising model universality class
for critical behavior.
In particular,
owing to such a decay of $\Delta$
the ground--state energy per site
in the vicinity of $\Omega^\star$
has the form
\begin{eqnarray}
\label{3.07}
e_0=-\int_0^\infty
{\mbox{d}}EE^2R(E^2)
\nonumber\\
=-\frac{1}{p\pi}
\int_{\sqrt{\epsilon^2}}^{\sqrt{a_2}}
{\mbox{d}}EE^2\frac{f(E^2)}{\sqrt{E^2-\epsilon^2}}
\nonumber\\
+{\mbox{analytical with respect to}}\;\epsilon^2\;
{\mbox{terms}}.
\end{eqnarray}
Here the first term is a contribution of the lowest energy band
and the explicit expression for $f(E^2)$
is not important for the analysis of nonanalytical behavior as $\epsilon\to 0$.
The first term in (\ref{3.07}) is proportional to
$\epsilon^2\ln\epsilon$
and as a result the zero--temperature dependence of
$m^z$ and $\chi^z$ on $\Omega$
contains the nonanalytical terms
$\left(\Omega-\Omega^\star\right)\ln\vert\Omega-\Omega^\star\vert$
and
$\ln\vert\Omega-\Omega^\star\vert$,
respectively.

Let us turn to the case $p=2$
with $\Delta\Omega=\sqrt{\vert I_1I_2\vert}$
when we have three critical fields
$\Omega^\star
=\left\{\pm\sqrt{2\vert I_1I_2\vert},0\right\}$
and consider the behavior of $\Delta(\Omega)$
in the vicinity of $\Omega^\star=0$,
i.e., as $\Omega\to 0$.
From Eq. (\ref{3.06}) one finds
that $\Delta^2(\Omega)\sim\epsilon^4$.
Repeating the calculation of the ground--state energy
(see Eq. (\ref{3.07}))
for such a decay of $\Delta$
one finds that $e_0$ contains the term
$\epsilon^4\ln\epsilon$
and hence the system exhibits the fourth--order
(in Ehrenfest's sense)
quantum phase transition
at $\Omega^\star=0$
which is
characterized by a logarithmic divergence
of the second derivative of the susceptibility
$\frac{\partial^2\chi^z}{\partial \Omega^2}$.
(For an example of a fourth--order thermal phase transition
see Ref. \onlinecite{027}.)
For $p=3$ the dependence
$\Delta(\Omega)\sim\epsilon^2$
(see Figs. \ref{fr01}b -- \ref{fr01}d)
may occur for the sets of parameters at the boundaries
between different regions in Fig. \ref{fr02}
(e.g., at the points b, c, e).
Such systems again show
the fourth--order quantum phase transition behavior
while approaching the corresponding critical fields.

\begin{figure}[t]
\vspace{-15mm}
\epsfxsize=10.0cm
\center{\epsffile{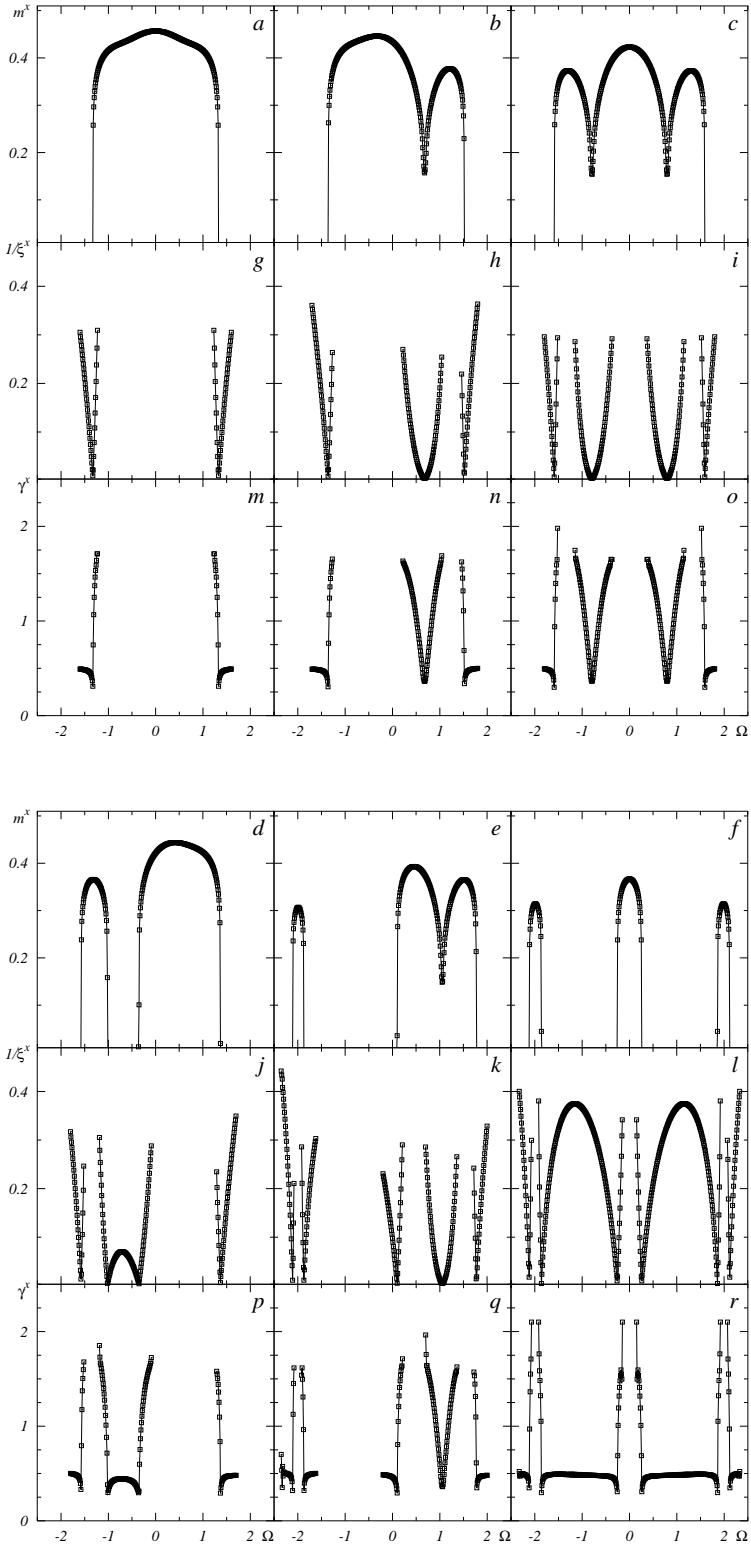}}
\caption
{The ground--state $x$--magnetization
$m^x=\frac{1}{3}\left(\vert m_1^x\vert+\vert m_2^x\vert+\vert m_3^x\vert\right)$
(a, b, c, d, e, f),
inverse correlation length $\frac{1}{\xi^x}$ (g, h, i, j, k, l),
and $\gamma^x$ (m, n, o, p, q, r)
for the transverse Ising chain of period 3
with
$\vert I_1\vert=\vert I_2\vert=\vert I_3\vert=1$,
$\Omega_{1,2,3}
=\Omega+\Delta\Omega_{1,2,3}$,
$\Delta\Omega_1+\Delta\Omega_2+\Delta\Omega_3=0$,
$\Delta\Omega_1=1$,
$\Delta\Omega_2=0$
($\Omega^\star
\approx\left\{\pm 1.325 \right\}$) (a, g, m),
$\Delta\Omega_1=1$,
$\Delta\Omega_2\approx 0.292$
($\Omega^\star
\approx \left\{ -1.355, 0.678, 1.513 \right\}$) (b, h, n),
$\Delta\Omega_1=\left(\frac{27}{4}\right)^\frac{1}{6}\approx 1.374$,
$\Delta\Omega_2=0$
($\Omega^\star
\approx
\left\{\pm 0.794,\pm 1.587\right\}$) (c, i, o),
$\Delta\Omega_1=\left(\frac{27}{4}\right)^\frac{1}{6}\approx 1.374$,
$\Delta\Omega_2=-1$
($\Omega^\star
\approx
\left\{-1.574, -1.020, -0.348, 1.367\right\}$) (d, j, p),
$\Delta\Omega_1=2$,
$\Delta\Omega_2\approx -1.575$
($\Omega^\star
\approx
\left\{- 2.107, -1.874, 0.102, 1.054, 1.772\right\}$) (e, k, q),
$\Delta\Omega_1=2$,
$\Delta\Omega_2=0$
($\Omega^\star
\approx
\left\{\pm 0.254, \pm 1.861, \pm 2.115 \right\}$) (f, l, r).
Connecting curves are guides to the eye.
The taken sets of parameters are in correspondence with the points
a, b, c, d, e, f
in Fig. \ref{fr02}.
\label{fr05}}
\end{figure}

\begin{figure}[t]
\epsfxsize=16.0cm
\center{\epsffile{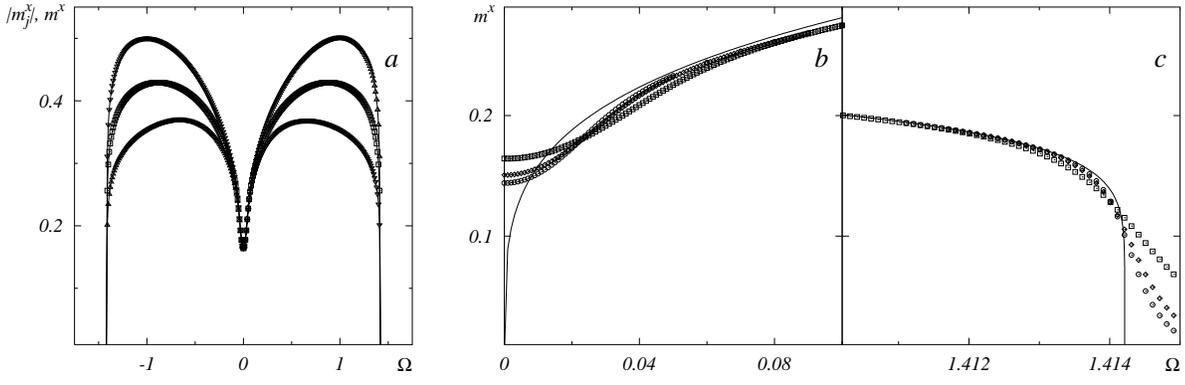}}
\caption
{The ground--state sublattice $x$--magnetizations
$\vert m_j^x\vert$, $j=1,2$
(triangles)
and
$m^x=\frac{1}{2}\left(\vert m_1^x\vert+\vert m_2^x\vert\right)$
(squares, diamonds, circles)
for the transverse Ising chain of period 2
with
$\vert I_1\vert=\vert I_2\vert=1$,
$\Omega_{1,2}=\Omega\pm 1$.
Connecting curves in panel a are guides to the eye.
We also report the results for the dependence $m^x$ vs. $\Omega$
in the vicinity of the critical points
$\Omega^\star=0$ (b)
and
$\Omega^\star=\sqrt{2}$ (c)
(squares, diamonds, circles
correspond to data for $N=2000,\;4000,\;5400$,
respectively,
solid lines represent dependences
$\sim\Omega^\frac{1}{4}$ (b)
and
$\sim\left(\sqrt{2}-\Omega\right)^\frac{1}{8}$ (c)).
\label{fr06}}
\end{figure}

To discuss further the quantum phases
which occur as the transverse field varies
we examine the spin correlation functions
$\langle s_n^\alpha s_{n+l}^\alpha\rangle$.
Unfortunately,
we cannot obtain the spin correlation functions
of a regularly alternating transverse Ising chain
using the continued fraction approach
which is restricted to the quantities
that can be expressed through the density of states (\ref{2.04}).
However,
the spin correlation functions
can be determined numerically
(see, for example, Ref. \onlinecite{028})
for rather long chains of several thousand sites.
Knowing $\langle s_n^\alpha s_{n+l}^\alpha\rangle$
we can obtain the on-site magnetization
${m_n^\alpha}^2
=\mbox{lim}_{r\to\infty}\langle s_n^\alpha s_{n+rp}^\alpha\rangle$.
Assuming that
$\langle s_n^\alpha s_{n+rp}^\alpha\rangle
-\langle s_n^\alpha\rangle \langle s_{n+rp}^\alpha\rangle$
decays as
$\left(rp\right)^{-\gamma^\alpha}\exp\left(-\frac{rp}{\xi^\alpha}\right)$
if $r\to\infty$
we can also find the correlation length
$\xi^\alpha$
and the power--law exponent
$\gamma^\alpha$.
In our calculations
for $p=3$
we consider chains
with $N=2100$,
take $n=500$, $rp=999$
to determine $\vert m_j^x\vert$
and
$rp=60,\ldots,360$
to determine $\xi^x$ and $\gamma^x$.
Our findings are collected in Fig. \ref{fr05}.
To illustrate the critical behavior of the order parameter in detail
we also consider a chain of period 2
with $\vert I_1\vert= \vert I_2\vert=1$,
$\Omega_{1,2}=\Omega\pm 1$
taking $N=2000, \; 4000,\; 5400$,
$n=\frac{N}{4}$,
$rp=\frac{N}{2}$.
The zero--temperature dependences
$\vert m^x_j\vert$ vs. $\Omega$ for this chain
are reported in Fig. \ref{fr06}.

As can be seen from Figs. \ref{fr05}, \ref{fr06}
the behavior of the Ising magnetization $m^x$
(which plays the role of the order parameter)
indicates the different phases
(Ising phase if $m^x\ne 0$
or paramagnetic phase if $m^x=0$)
and the phase transitions between them.
For a set of parameters which yields weak singularities
(i.e., $m^x=0$ in the Ising phase,
see Figs. \ref{fr05}b, \ref{fr05}c, \ref{fr05}e, \ref{fr06}a, \ref{fr06}b)
the finite--size effects are strong
and the finite--chain result for $x$--magnetization
tends to zero very slowly
with increasing $N$
(compare data for different $N$ in Fig. \ref{fr06}b).
For the second--order (fourth--order)
quantum phase transition points
the critical behavior is given by
$m^x\sim\vert \Omega^\star-\Omega\vert^\beta$
with $\beta=\frac{1}{8}$ ($\beta=\frac{1}{4}$)
(compare Figs. \ref{fr06}c and \ref{fr06}b).
The appearance/disappearance of the Ising magnetization
is accompanied
by a divergence of the correlation length
$\xi^x=\vert\Omega-\Omega^\star\vert^{-\nu}$
with $\nu=1$ ($\nu=2$)
for the second--order (fourth--order) quantum phase transition points
(Figs. \ref{fr05}g -- \ref{fr05}l).
Taking into account
the values of the exponent
characterizing the energy gap behavior
we conclude
that the relaxation time scales like the first power of the correlation length,
i.e., the dynamic exponent $z=1$,
for both the second--order and the fourth--order quantum phase transitions.
At $\Omega=\Omega^\star$
the $xx$ spin correlation functions show power--law decay with the exponent $\gamma^x=\frac{1}{4}$
for both the second--order and the fourth--order quantum phase transitions
(Figs. \ref{fr05}m -- \ref{fr05}r).
Finally,
the results for spin correlation functions
at special values of the transverse field $\Omega$,
i.e., when one on--site field equals zero,
coincide with the analytical predictions
obtained
using the 3--site cluster Hamiltonian eigenvectors
(for a chain of period 2
the corresponding calculations are given in Ref. \onlinecite{029}).
For example,
for the chain of period 3
with $I_1=I_2=I_3=1$,
$\Delta\Omega_1=1$,
$\Delta\Omega_2=0$,
$\Delta\Omega_3=-1$
at $\Omega=-1$
we have found
$\vert m_1^x\vert =\frac{1}{2},
\;\;\;
\vert m_2^x\vert \approx 0.417,
\;\;\;
\vert m_3^x\vert \approx 0.331$
(see Fig. \ref{fr05}a)
whereas
for such a chain
with
$\Delta\Omega_1=2$,
$\Delta\Omega_2=0$,
$\Delta\Omega_3=-2$
at $\Omega=-2$
we have found
$\vert m_1^x\vert =\frac{1}{2},
\;\;\;
\vert m_2^x\vert \approx 0.267,
\;\;\;
\vert m_3^x\vert \approx 0.175$
(see Fig. \ref{fr05}f).
It should be noted
that the Ising magnetization
at the sites with zero transverse fields has its maximal value
$\frac{1}{2}$.
Probably the most spectacular feature
of the Ising chain with regularly alternating transverse field
is the reentrant behavior with varying $\Omega$
nicely seen in Figs. \ref{fr05}d -- \ref{fr05}f.
The appearance of the paramagnetic phase
at intermediate values of the transverse field
when $x$--magnetization is zero
and $z$--magnetization is almost constant
can be associated with the following classical picture.
Assume, for example,
$p=2$ and $\Omega=0$;
then owing to the regularly varying on--site transverse fields
$\pm\Delta\Omega$
with large $\Delta\Omega$
all on--site magnetizations
are directed in $\pm z$--direction in the spin space.
Naturally,
this picture may play only an auxiliary role
for the considered quantum systems.

Finally
we note
that our results
are in agreement with the scaling relations
in the theory of conventional (temperature--driven) phase transitions\cite{030}.
Thus, the quantum phase transition in dimension $d=1$
corresponds to the thermal phase transition in dimension $d+z=2$,
the exponent $\nu$ which characterizes the divergence of the correlation length
$\xi\sim\vert T-T_c\vert^{-\nu}$
characterizes
the decay of the energy gap
$\Delta\sim \vert \Omega-\Omega^\star\vert^{\nu}$,
the exponent $\alpha$ characterizing the divergence of the specific heat
$c\sim\vert T-T_c\vert^{-\alpha}$
characterizes the divergence of the transverse susceptibility
$\chi^z\sim\vert\Omega-\Omega^\star\vert^{-\alpha}$.
Moreover,
a number of scaling relations
(which do not account for logarithmic divergences)
hold.
For example,
$2-\alpha=d\nu$.
Substituting $d=2$, $\nu=1$ one gets $\alpha=0$,
i.e., only a logarithmic divergence
in the dependence $\chi^z$ vs. $\Omega$,
whereas for more rapidly decaying energy gap when $\nu=2$
one finds
$\alpha=-2$,
i.e., $\chi^z$ does not diverge at $\Omega^\star$
(and only its second derivative exhibits a logarithmic peculiarity).

\subsection{Temperature behavior of the specific heat}

\begin{figure}[t]
\epsfxsize=16.0cm
\center{\epsffile{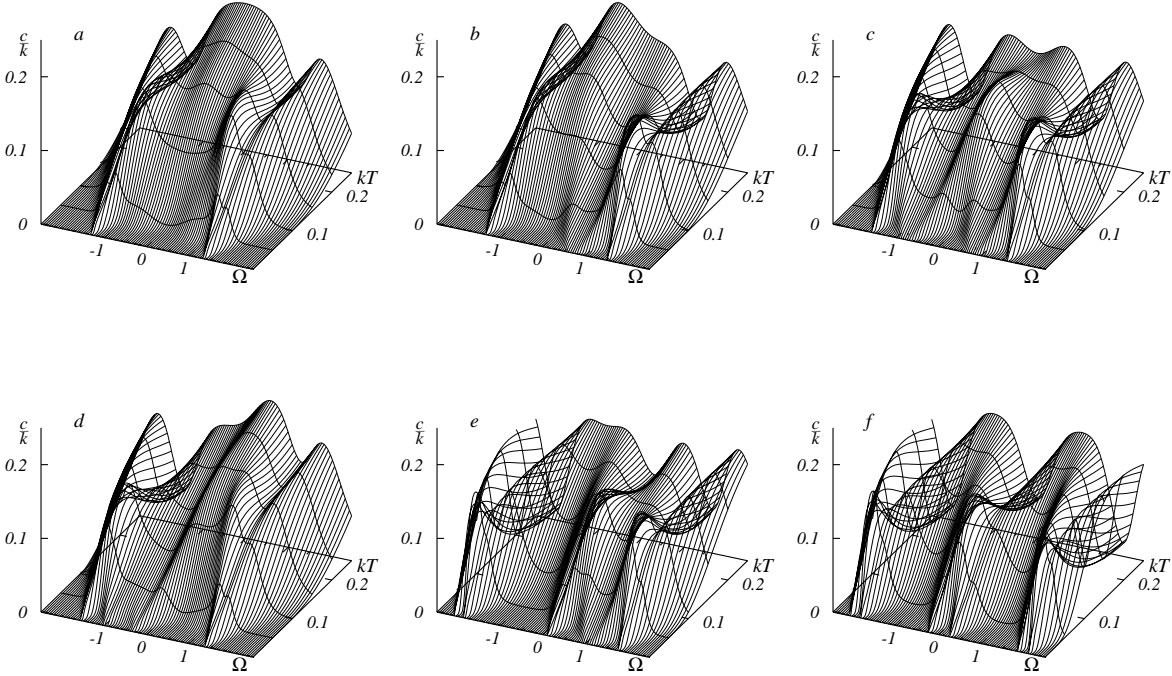}}
\caption
{The low--temperature behavior of the specific heat
for the transverse Ising chain of period 3
with $\vert I_1\vert=\vert I_2\vert=\vert I_3\vert=1$,
$\Omega_{1,2,3}=\Omega+\Delta\Omega_{1,2,3}$,
$\Delta\Omega_1+\Delta\Omega_2+\Delta\Omega_3=0$,
$\Delta\Omega_1=1$, $\Delta\Omega_2=0$ (a),
$\Delta\Omega_1=1$, $\Delta\Omega_2\approx 0.292$ (b),
$\Delta\Omega_1\approx 1.374$, $\Delta\Omega_2=0$ (c),
$\Delta\Omega_1\approx 1.374$, $\Delta\Omega_2=-1$ (d),
$\Delta\Omega_1=2$, $\Delta\Omega_2\approx -1.575$ (e),
$\Delta\Omega_1=2$, $\Delta\Omega_2=0$ (f).
\label{fr07}}
\end{figure}

We turn to a discussion of the effects of regular alternation
on the temperature dependence
of the specific heat.
The low--temperature behavior of this quantity
is determined by the fact
whether the system is gapped or gapless.
Thus,
the zero--energy excitations
immediately produce a linear dependence
of the specific heat on temperature.
As a result
the low--temperature behavior of the specific heat
indicates the quantum phase transition points
that can be seen in Fig. \ref{fr07}
in complete agreement with the outcomes which follow from the behavior
of the $xx$ spin correlation functions
shown in Fig. \ref{fr05}.
Moreover,
we notice that the regular alternation may produce
many--peak structure of the temperature profiles of the specific heat
(Fig. \ref{fr07}).

\section{Anisotropic $XY$ chain without field. Spin--Peierls dimerization}
\label{s4}
\setcounter{equation}{0}

As a byproduct of the study
of regularly alternating transverse Ising chains
we obtain the thermodynamic quantities
of regularly alternating anisotropic $XY$ chains without field (\ref{1.03}).
Really,
using the unitary transformations
discussed in the end of Section \ref{s1}
we can state
that the Helmholtz free energy
of the regularly alternating anisotropic $XY$ chain without field
(\ref{1.03})
defined by a sequence of parameters
$I_1^xI_1^yI_2^xI_2^y\ldots I_p^xI_p^y
I_1^xI_1^yI_2^xI_2^y\ldots I_p^xI_p^y
\ldots$
is given by Eq. (\ref{2.05})
with $R(E^2)$ (\ref{2.07})
and the diagonal Green functions involved into Eq. (\ref{2.07})
are determined as follows
\begin{eqnarray}
\label{4.01}
G_{nn}
=\frac{1}
{E^2-{I^x_{n-1}}^2-{I^y_{n}}^2-\Delta_n^--\Delta_n^+},
\\
\Delta_n^-
=\frac{{I^y_{n-2}}^2{I^x_{n-1}}^2}
{E^2-{I^x_{n-3}}^2-{I^y_{n-2}}^2
-\frac{{I^y_{n-4}}^2{I^x_{n-3}}^2}
{E^2-{I^x_{n-5}}^2-{I^y_{n-4}}^2-_{\ddots}}},
\nonumber\\
\Delta_n^+
=\frac{{I^y_{n}}^2{I^x_{n+1}}^2}
{E^2-{I^x_{n+1}}^2-{I^y_{n+2}}^2
-\frac{{I^y_{n+2}}^2{I^x_{n+3}}^2}
{E^2-{I^x_{n+3}}^2-{I^y_{n+4}}^2-_{\ddots}}}.
\nonumber
\end{eqnarray}

Moreover,
we may use the obtained densities of states (\ref{2.09}), (\ref{2.10}), (\ref{2.11})
to find the thermodynamic quantities
of some regularly alternating anisotropic $XY$ chains.
Thus,
the anisotropic $XY$ chain of period 2
is unitary equivalent
to two different transverse Ising chains
both of period 1
and as a result
\begin{eqnarray}
\label{4.02}
R(E^2)
=
\left\{
\begin{array}{ll}
\frac{1}{2\pi}\frac{1}{\sqrt{{\cal{A}}_{xy}(E^2)}}, &
{\mbox{if}} \;\;\; {\cal{A}}_{xy}(E^2)>0\\
0, &
{\mbox{otherwise}}
\end{array}
\right.
\nonumber\\
+
\left\{
\begin{array}{ll}
\frac{1}{2\pi}\frac{1}{\sqrt{{\cal{A}}_{yx}(E^2)}}, &
{\mbox{if}} \;\;\; {\cal{A}}_{yx}(E^2)>0,\\
0, &
{\mbox{otherwise}},
\end{array}
\right.
\\
{\cal{A}}_{\alpha\beta}(E^2)
=
-\left(
E^2-\left(I_1^\alpha-I_2^\beta\right)^2
\right)
\left(
E^2-\left(I_1^\alpha+I_2^\beta\right)^2
\right).
\nonumber
\end{eqnarray}
(Note,
that for the isotropic case
$I_1^x=I_1^y=I_1$,
$I_2^x=I_2^y=I_2$,
Eq. (\ref{4.02}) yields the result obtained in Ref. \onlinecite{024}
(Eqs. (9) -- (11) of that paper);
in the anisotropic case Eq. (\ref{4.02})
agrees with the result reported in Ref. \onlinecite{010}.)
The anisotropic $XY$ chain of period 3
after performing the above mentioned unitary transformations
is equivalent
to two identical transverse Ising chains of period 3
and, therefore,
$R(E^2)$
is given by Eqs. (\ref{2.09}), (\ref{2.11})
after the substitution
$\Omega_1\to I_1^y$,
$I_1\to I_2^x$,
$\Omega_2\to I_3^y$,
$I_2\to I_1^x$,
$\Omega_3\to I_2^y$,
$I_3\to I_3^x$.
Let us also note
that the anisotropic $XY$ chains of period 4 (6)
are unitary equivalent to
two different transverse Ising chains of period 2 (3)
and hence
after simple substitutions
Eqs. (\ref{2.09}), (\ref{2.10}) (Eqs. (\ref{2.09}), (\ref{2.11}))
yield the thermodynamic properties of such chains.

\begin{figure}[t]
\epsfxsize=11.5cm
\center{\epsffile{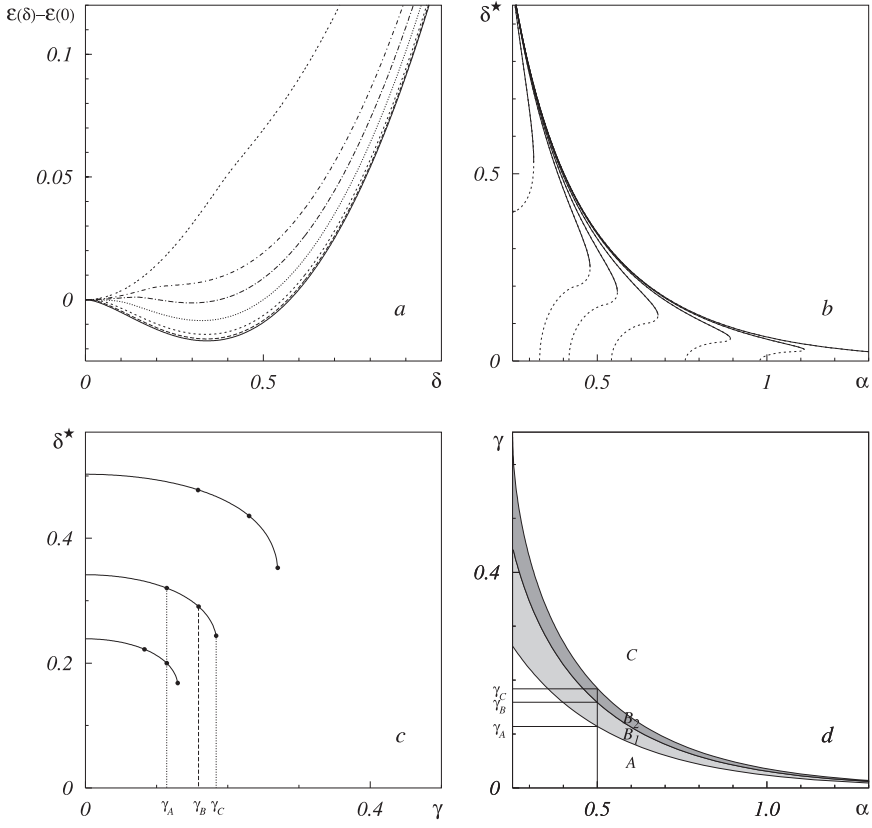}}
\caption
{The total ground--state energy per site $\cal{E}(\delta)$
vs.
dimerization parameter $\delta$
($\alpha=0.5$,
from bottom to top $\gamma=0,\;0.025,\;0.05,\;0.1,\;0.15,\;0.2,\;0.4$)
(a),
the dimerization parameter $\delta^\star$ vs. $\alpha$
(from right to left $\gamma=0,\;0.025,\;0.05,\;0.1,\;0.15,\;0.2,\;0.4$)
(b),
the dimerization parameter $\delta^\star$ vs. $\gamma$
(from top to bottom $\alpha=0.4,\;0.5,\;0.6$;
the meaning of the characteristic values of the anisotropy parameter
$\gamma_A$,
$\gamma_B$,
$\gamma_C$
(denoted for $\alpha=0.5$)
is explained in the main text)
(c),
and
the phase diagram in the plane
$\alpha$ -- $\gamma$
(in the region A (C) the dimerized (uniform) phase occurs;
in the regions B$_1$, B$_2$ both phases are possible
although in the region B$_1$ (B$_2$) the dimerized (uniform) phase is favorable)
(d).
\label{fr08}}
\end{figure}

Let us use the ground--state energy per site
$e_0=-2\int_0^\infty{\mbox{d}}E E^2 R(E^2)$
of the anisotropic $XY$ chain of period 2
to examine
the effects of the exchange interaction anisotropy
on the spin--Peierls dimerization
inherent in the isotropic $XY$ chain\cite{031,032}.
For this purpose we assume
in (\ref{4.02})
$I_1^x=(1+\delta)(1+\gamma)$,
$I_1^y=(1+\delta)(1-\gamma)$,
$I_2^x=(1-\delta)(1+\gamma)$,
$I_2^y=(1-\delta)(1-\gamma)$
where $0\le\delta<1$ is the dimerization parameter
and $0\le \gamma\le 1$ is the exchange interaction anisotropy parameter.
We consider
the total energy per site ${\cal{E}}(\delta)$
and its dependence on $\delta$.
${\cal{E}}(\delta)$ consists of the magnetic part $e_0(\delta)$
and the elastic part $\alpha\delta^2$.
Let us recall that in the isotropic limit $\gamma=0$
the total energy ${\cal{E}}(\delta)$
exhibits a minimum at a nonzero value of the dimerization parameter
$\delta^\star\ne 0$
which is a manifestation of lattice instability
with respect to spin--Peierls dimerization\cite{031}.
In the other limiting case $\gamma=1$
the magnetic energy does not depend on $\delta$
and hence the uniform lattice is stable.
In Fig. \ref{fr08}a
one can see how the behavior
of ${\cal{E}}(\delta)$ vs. $\delta$
varies as $\gamma$ increases from 0 to 0.4
for $\alpha=0.5$.
At $\gamma=0$
the total energy ${\cal{E}}(\delta)$ exhibits a minimum
at a nonzero value of dimerization parameter $\delta^\star\ne 0$.
As $\gamma$ increases the dependence remains qualitatively the same
with, however, slightly decreasing value of $\delta^\star$
(see Figs. \ref{fr08}b and \ref{fr08}c).
At a certain value of anisotropy parameter $\gamma_A$
an additional minimum at $\delta=0$ appears.
Both minima are separated by a maximum,
and the minimum at $\delta^\star\ne 0$ remains the deeper one.
At the value of $\gamma_B$ ($>\gamma_A$)
the minima have the same depth
and with further increase of $\gamma$ the minimum at $\delta=0$
becomes the deeper one.
If $\gamma$ exceeds $\gamma_C$ ($>\gamma_B$)
the minimum for a nonzero dimerization parameter disappears.
In Fig. \ref{fr08}b
one can see the behavior of $\delta^\star$
as $\gamma$ varies from 0 to 0.4
for different $\alpha$s
(solid curves; the dashed curves show the behavior of the maximum
in the dependence ${\cal{E}}(\delta)$ vs. $\delta$);
in Fig \ref{fr08}c
one can see the dependence of $\delta^\star$ on $\gamma$
for $\alpha=0.4,\;0.5,\;0.6$.
The bold dots in the curves in this panel
correspond to the characteristic values of the anisotropy parameter
$\gamma_A<\gamma_B<\gamma_C$
discussed above.
The effect of the anisotropy on the spin--Peierls dimerized phase
occurs according to the first--order phase transition scenario.
The corresponding phase diagram is shown in Fig. \ref{fr08}d
where we indicate
the region of stability of the dimerized (A) and uniform (C) phases
as well as the metastable region (regions B$_1$ and B$_2$)
where both phases may coexist.

It is interesting to compare
the described effects of the exchange interaction anisotropy
on the spin--Peierls dimerized phase
with the effects of the transverse field
on the spin--Peierls dimerized phase\cite{032,024}.
Similarly to the anisotropy $\gamma$
the transverse field $\Omega$
destroys the dimerized phase
according to a first--order phase transition scenario.
However, the value of the dimerization parameter $\delta^\star$
remains unchanged as $\Omega$ increases up to $\Omega_C$.

\section{Concluding remarks}
\label{s5}
\setcounter{equation}{0}

In this work we have analyzed in some detail
the ground--state and thermodynamic properties
of regularly alternating spin--$\frac{1}{2}$ transverse Ising chains
and anisotropic $XY$ chains without field.
Due to the Jordan--Wigner mapping
and the continued fraction approach
we can calculate the thermodynamic quantities rigorously analytically.
For certain values of parameters
we can also calculate the ground--state spin correlation functions.
For other values of parameters
we have calculated the spin correlation functions numerically
for long chains consisting of a few thousand sites.
We have shown how the ground--state properties
of regularly alternating classical transverse Ising/$XX$ chains
can be examined.
The main new results obtained are as follows.
Firstly,
we have examined the effects of regular alternation
on quantum phases and quantum phase transitions
in the transverse Ising chain.
Owing to regularly alternating parameters
the number of quantum phase transition may increase
(but never exceeds $2p$
where $p$ is the period of alternation),
the critical behavior remains as in the uniform chain,
however, a weaker singularity may also appear.
Secondly,
we have demonstrated how
the plateaus in the ground--state magnetization curves
for the classical regularly alternating transverse Ising/$XX$ chains
may emerge.
Thirdly,
we have shown
how
the exchange interaction anisotropy destroys the spin--Peierls dimerization
inherent in the spin--$\frac{1}{2}$ isotropic $XY$ chain.
The performed study provides a set of reference results
which may be useful for understanding more complicated quantum spin chains.

\section*{Acknowledgments}

The present study was partly supported by the DFG
over
a few past years
a number of times
(projects
436 UKR 17/7/01,
436 UKR 17/1/02,
436 UKR 17/17/03).
O. D. acknowledges the kind hospitality of the Magdeburg University
in the autumn of 2003
when the paper was completed.
The paper was partly presented
at the 19th General Conference of the EPS Condensed Matter Division
held jointly with
CMMP 2002 -- Condensed Matter and Materials Physics
(Brighton, UK, 2002),
at the 28th Conference of the Middle European Cooperation in Statistical Physics
(Saarbr\"{u}cken, Germany, 2003)
and
at the International Workshop and Seminar on Quantum Phase Transitions
(Dresden, Germany, 2003).
O. D. expresses appreciation to
the Institute of Physics
for the support in participation.
O. D. and O. Z. thank the organizers of the MECO conference
for the support.
O. D. is grateful to the Max--Planck--Institut f\"{u}r Physik komplexer Systeme
for the hospitality in Dresden.

\clearpage

\input epsf

\input epsf

\input epsf

\input epsf

\input epsf

\input epsf

\input epsf

\input epsf

\end{document}